\documentclass[12pt]{article}
\usepackage{amssymb,amsmath,latexsym,graphicx,tikz,setspace,wasysym,
float,subfig}
\usepackage{hyperref}
\pdfoutput=1

\newcommand{\beq}{\begin{equation}}
\newcommand{\be}{\begin{equation}}
\newcommand{\ee}{\end{equation}}
\newcommand{\bea}{\begin{eqnarray}}
\newcommand{\eea}{\end{eqnarray}}

\newcommand{\pa}{\partial}

\newcommand{\nn}{\nonumber}

\newcommand{\mc}{\mathcal}

\newcommand{\zb}{\bar{z}}
\newcommand{\tb}{\bar{t}}

\numberwithin{equation}{section}
\onehalfspace
\begin{document}

\begin{titlepage}
\hbox to \hsize{\hspace*{0 cm}\hbox{\tt }\hss
   \hbox{\small{\tt }}}

\vspace{1 cm}

\centerline{\bf \Large Twist-nontwist correlators in $M^N/S_N$ orbifold CFTs}

\vspace{.6cm}

\vspace{1 cm}
 \centerline{\large Benjamin A. Burrington$^\ddagger$\footnote[1]{bburington@troy.edu}\,, Amanda W. Peet$^\dagger$\footnote[2]{amanda.peet@utoronto.ca}\,, and  Ida G. Zadeh$^\dagger$\footnote[3]{ghazvini@physics.utoronto.ca}}

\vspace{0.3cm}
\centerline{\it ${}^\ddagger$Department of Chemistry and Physics,
%}
%\centerline{\it
Troy University,
%}
%\centerline{\it
Troy, Alabama, USA 36082.}

\vspace{0.3cm}
\centerline{\it ${}^\dagger$Department of Physics,%}
%\centerline{\it
University of Toronto,%}
%\centerline{\it
Toronto, Ontario, Canada M5S 1A7. }

\vspace{0.3 cm}

\begin{abstract}

We consider general 2D orbifold CFTs of the form $M^N/S_N$, with $M$ a target space manifold and $S_N$ the symmetric group, and generalize the Lunin-Mathur covering space technique in two ways.  First, we consider excitations of twist operators by modes of fields that are not twisted by that operator, and show how to account for these excitations when computing correlation functions in the covering space. Second, we consider non-twist sector operators and show how to include the effects of these insertions in the covering space.  We work two examples, one using a simple bosonic CFT, and one using the D1-D5 CFT at the orbifold point.  We show that the resulting correlators have the correct form for a 2D CFT.

\end{abstract}

\end{titlepage}

%--------+---------+---------+---------+---------+---------+---------+---------+
\tableofcontents

%--------+---------+---------+---------+---------+---------+---------+---------+

\section{Introduction}\label{introduction}

The holographic connection between gravitational physics and lower dimensional field theories provided by AdS/CFT  has become a fundamental tool to study both sides of the duality (see \cite{Aharony:1999ti} for reviews).  Of primary importance is the application of field theory techniques to elucidate black hole thermodynamics, and to obtain quantitative descriptions of microstates for gravitational systems.

One context where AdS/CFT has been used to great advantage is the D1-D5 system \cite{Vafa:1995bm,Maldacena:1997re,Dijkgraaf:1998gf,Jevicki:1998bm,Giveon:1998ns,Seiberg:1999xz,Gava:2001ne}  (see also \cite{David:2002wn,Bena:2007kg} for reviews).  The near horizon geometry of the D1-D5 system is AdS$_3 \times S ^3 \times M$ where $M$ is either $T^4$ or $K3$.  This geometry has been studied heavily in the literature, and forms the basis for the fuzzball proposal for black hole thermodynamics \cite{Lunin:2001jy,Lunin:2002qf,Mathur:2002ie}, for reviews see \cite{Mathur:2005zp,Mathur:2005ai,Balasubramanian:2008da,Skenderis:2008qn,Chowdhury:2010ct}.  Both gravitational and 2D CFT techniques have proven very useful in the analysis \cite{deBoer:1998ip, Seiberg:1999xz, Larsen:1999uk, David:1999ec, Lunin:2002fw, Gava:2002xb, Rychkov:2005ji, Avery:2009tu, Avery:2010er, Avery:2010hs, Avery:2010qw, Giusto:2012yz, Lunin:2012new}.

The CFT dual is believed to posses a position in moduli space, known as the orbifold point, where it is an $\mathcal{N}=(4,4)$ supersymmetric sigma model with target space $M^N/S_N$ \cite{Vafa:1995bm,Dijkgraaf:1998gf}.  Here, $M$ is either $K3$ or $T^4$, and $S_N$ is the symmetric group.  The target space $M^N$ would simply be $N$ copies of the CFT with target space $M$, and the $S_N$ symmetry acts by permutating the copies.  The fact that the theory admits this orbifold point is why we are interested in studying $S_N$ orbifold CFTs.

This motivates us to study 2D orbifold CFTs \cite{Dixon:1985jw,Dixon:1986qv,Dijkgraaf:1989hb}, with particular attention paid to symmetric group orbifolds \cite{Arutyunov:1997gt,Arutyunov:1997gi,Jevicki:1998bm,Lunin:2000yv,Lunin:2001pw,Pakman:2009ab,Pakman:2009zz,Pakman:2009mi}. Orbifolds provide models that inherit much of the parent theory's simplicity.  Orbifolds of free theories are particularly simple to deal with; the orbifold point D1-D5 CFT belongs to this class.   The added complication arises from the new sectors of the Hilbert space, the twisted sector states.  Such states are configurations that return to themselves up to an application of the orbifold symmetry, i.e. they are a new set of allowed boundary conditions when the theory is put on a cylinder.  New operators must be associated with these new states, according to the state-operator mapping.  However, it seems at first glance that these operators are rather implicitly defined, only given by a new set of boundary conditions on the states.

It is exactly this concern that Lunin and Mathur \cite{Lunin:2000yv,Lunin:2001pw} addressed for symmetric group $S_N$ orbifolds.  The basic idea in \cite{Lunin:2000yv} is to ``untwist'' the boundary conditions by using a locally conformal map.  We refer to the space where the CFT is originally defined as the base space, and the space to which it is mapped as the covering space.  The covering space is a multiple cover of the base space.  The purpose of the map is to make the fields in the covering space have simple boundary conditions, with the complications of boundary conditions in the base space being absorbed into the map between the two spaces.  In \cite{Lunin:2001pw}, the analysis was extended to the case where the CFT contains fermions, concentrating on the case for the orbifold point of the D1-D5 CFT.  Including fermions slightly complicates the picture in the covering space because of the conformal transformation properties of the fermions.  Further, \cite{Lunin:2001pw} discussed how to excite the twist sector operators with modes of fields that are twisted.  They concentrated on excitations involving symmetry currents in the theory, and showed how this operation is lifted to the covering surface.  The twisted currents have fractional modes with very low conformal dimension, a fact which they exploited in \cite{Lunin:2001pw} to construct super chiral primary operators in each twist sector.

The purpose of our work is to extend the Lunin-Mathur (LM) construction in two ways.  First, consider excitations of twist operators by modes that are not twisted by that operator.  It is relatively easy to construct the non-$S_N$-invariant operators because the operator being appended to the twist operator shares no OPE with it.  However, these added excitations can be twisted by other operators appearing in a correlator.  So, while the excitations are not twisted by the operator they excite, they may be twisted by other operators in the correlator.  We will explain how to account for these excitations in the covering space in the next section.

Our second extension is to find how to calculate correlators that contain both twist and non-twist sector fields.  For this, we note that the location of a non-twist operator will have several images in the covering space.  Each of these images has a concrete meaning in terms of the fields defined on the base space, and imply that summing over insertions at each image is the correct procedure.  For each of these extensions, we perform a sample calculation, and show that the result gives the correct form of a 2D CFT correlator in terms of the base space information.  We concentrate on 3-point functions in our examples, but the procedures can be applied to four point functions, as we will show in a companion work \cite{companion}.

The remainder of this work is organized as follows.  We summarize the Lunin-Mathur covering space technique in section \ref{lm}, setting up our generalization which we discuss in section \ref{gen}.  We illustrate our method with two examples.  In section \ref{ortho} we consider a free $X$ CFT factor appended to an arbitrary CFT.  We use the modes of the $X$ operator to excite a bare twist operator in non-twisted directions.  We show how to compute a 3-point correlator involving this operator, using the covering space, and show that it has the correct form for a 2D CFT 3-point function.  In section \ref{nontwist}, we consider the D1-D5 CFT near the orbifold point.  We use some excited versions of the super chiral primaries constructed in \cite{David:1999ec,Lunin:2001pw}, and add a non-twist sector excitation constructed from the bosons.  We show that, after summing over the images, the correlator is in fact of the correct form.  We conclude in section \ref{disc} with some interpretations of the method, and a discussion of applications.

%--------+---------+---------+---------+---------+---------+---------+---------+

\section{The Lunin-Mathur technique, and generalizations%
}
\subsection{Lunin-Mathur}\label{lm}

Here we will discuss the Lunin-Mathur technique \cite{Lunin:2000yv,Lunin:2001pw} (see also \cite{Avery:2010qw}
 for more explanation).  The LM technology was originally developed for bosonic theories in \cite{Lunin:2000yv} and later generalized to theories with fermions in \cite{Lunin:2001pw}.  In these works, particular attention was paid to the twist operators associated with cycles in the $S_N$ group, as these form a set of basic building blocks for $S_N$ (all group elements of $S_N$ can be written as products of cycles that do not share indices).  Excitations along directions twisted by the operator were also considered in \cite{Lunin:2001pw}, with the emphasis on how current operators act on the twist sector fields.  Here, we summarize the salient features of LM.

First, consider the r\^ole of twist fields.  Twist fields change the boundary conditions that the fundamental fields must satisfy.  In particular, they consider the effects of the non-$S_N$-invariant twist operators $\sigma_{(12\cdots n)}$, which twist the first $n$ copies of the CFT.  These act on the fields as
\be
\Phi_1 \rightarrow \Phi_2 \rightarrow \Phi_3 \rightarrow \cdots \rightarrow \Phi_n \rightarrow \Phi_1 \label{SNmap}
\ee
where by $\Phi$ we mean a field of arbitrary weight and statistics.  Here, and in what follows, we will suppress all other indices other than the copy index because the $S_N$ permutation leaves other indices unchanged.  The full $S_N$-invariant operator is generated by summing over the $S_N$ ``images'' of the non-$S_N$-invariant operators:
\be
\sigma_n=\frac{\sqrt{n(N-n)!}}{n(N-n)!\sqrt{N!}}\sum_{g} \sigma_{g(12\cdots n)g^{-1}}=\sqrt{\frac{n(N-n)!}{N!}} \sum_{c\in C[(12\cdots n)]} \sigma_{c}.
\ee
In the first equality we sum over all group elements $g$ of the symmetric group $S_N$.  In the second equality, we sum over the conjugacy class of $(123\cdots n)$, which we denote $C[(123\cdots n)]$.  This second sum just sums over all possible distinct $n$-cycles.

We are ultimately concerned with the evaluation of correlators of the form
\be
\langle \sigma_{n_1} \sigma_{n_2} \sigma_{n_3}\cdots\rangle. \label{corgenformini}
\ee
To do so, it would be sufficient to understand the correlators involving only the non-$S_N$-invariant twist operators e.g.
\be
\langle \sigma_{(1,2,3\cdots n_1)}\; \sigma_{(2,3,4, \cdots n_2+1)}\; \sigma_{(8,9,10\cdots n_3 +7)}\cdots\rangle. \label{corgenform}
\ee
because the correlator (\ref{corgenformini}) is just a sum of such terms.

Finding a way to represent the correlators (\ref{corgenform}) was the primary goal in \cite{Lunin:2000yv,Lunin:2001pw}.  The basic idea is to map the problem of multiple copies of fields with twisted boundary conditions to a problem of one copy of fields with normal periodic boundary conditions.  This is accomplished with a locally conformal map.  The base space we will parameterize with the complex variables $(z, \zb)$ and the covering space we will parameterize with the complex variables $(t, \tb)$.

Let us imagine that there are $s$ distinct indices involved in the twists in (\ref{corgenform}).  In this case, we will pay attention only to these $s$ copies of the fields: the other copies do not interact with these twists, and this part of the correlator factorizes.  Now focusing on only this set of $s$ copies, we consider an $s$-fold cover of the space.  In the covering space, we only have one copy of the fields, but because a generic point in the base space corresponds to multiple points in the covering space, we actually have multiple copies of the fields defined in the base space.  Thus, the map from the covering space to the base space induces the correct number of functions/fields in the base space.  For the time being we will restrict ourselves to bosonic fields $\Phi$, and will consider the extension to fermions later in this section.

Next, when circling a twist insertion in the base space the fields must map as (\ref{SNmap}).   Thus, starting with the field $\Phi_1$, and circling the insertion of $\sigma_{(1,2,3,\cdots n)}$, we find that the function does not come back to itself, but rather comes back to $\Phi_2$.  Thus, in the covering space, the contour must be open such that the single function $\Phi$ is different at the endpoints.  It must be that these endpoints in the covering space are mapped to the same point in the base space.  Further, we construct the map from the base space to the covering space such that there are distinguished points in the covering space \cite{Lunin:2000yv}.  These distinguished ``ramified'' points are where the map looks locally like
\be
z-z_0=b (t-t_0)^{n_i} + \cdots.
\ee
For each $n_i$ cycle twist insertion, we must have one such point.  Such a point in the covering space is where $n_i$ images of the base space come together, and it is these points that enforce the boundary conditions (\ref{SNmap}).  If we consider a contour around this point in the covering surface, it actually winds around a point in the base space $n_i$ times.  This point in the base space is the location of the $\sigma_{n_i}$ insertion.

Next, we note that any given point on the contour around the twist insertion must have $s$ total images.  It is clear that we have identified $n_i$ of these near the distinguished point in the covering space.  The other $s-n_i$ images must be near other locations in the covering space.  These points are isolated ``non-ramified'' points, and so going once around these points in the covering space correspond to going once around the twist insertion in the base space.  This works in the case that each twist insertion is a cycle: for products of cycles, we just take the coincidence limit of the considerations here.

To help think about the map, we restrict our attention in the base space to a patch that does not have any contours that go around the twist insertions: we call this patch the simply connected patch.  We then may take an arbitrary point in this patch, and consider one of its images in the cover cover.  We consider expanding this neighborhood until it fills the simply connected patch; we consider the expansion of the neighborhood in the covering space as well.  This defines one image of the simply connected patch on the covering surface.  We may do this with the other image points as well, and find all $s$ copies of the simply connected patch.

To each of these patches, we assign a function $\Phi_i$ associated with it.  To help identify these patches, we consider the periodicity when going around a twist insertion.  A given patch will have a certain number of points that are images of the location of twist operators.  If we consider those that are non-ramified, this gives the location of operators that do not twist the function defined by the patch in question.  This information should identify the patch uniquely.

\begin{figure}[ht!]
\centering
\subfloat[base space \label{mapdiaga}]{
    \includegraphics{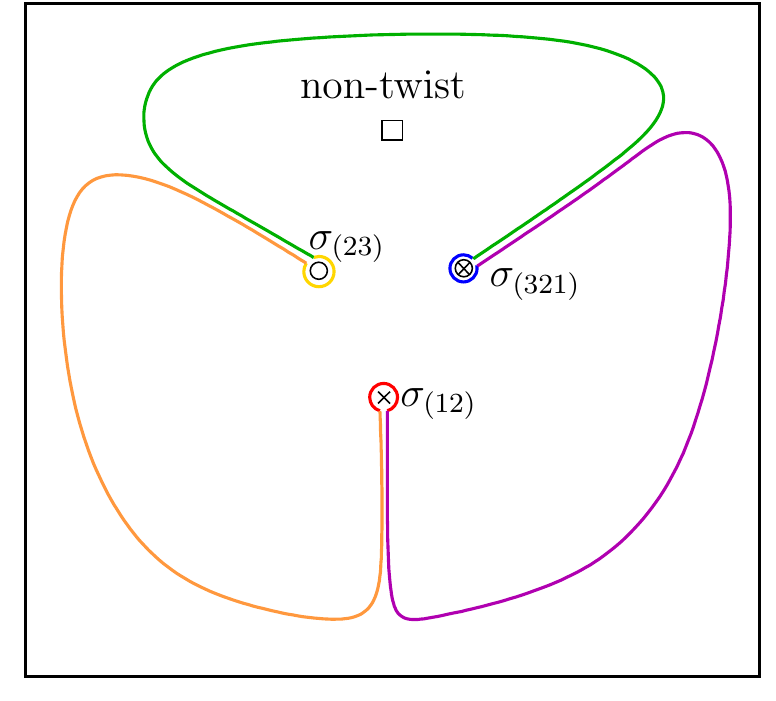}}
\subfloat[covering space \label{mapdiagb}]{
   \includegraphics{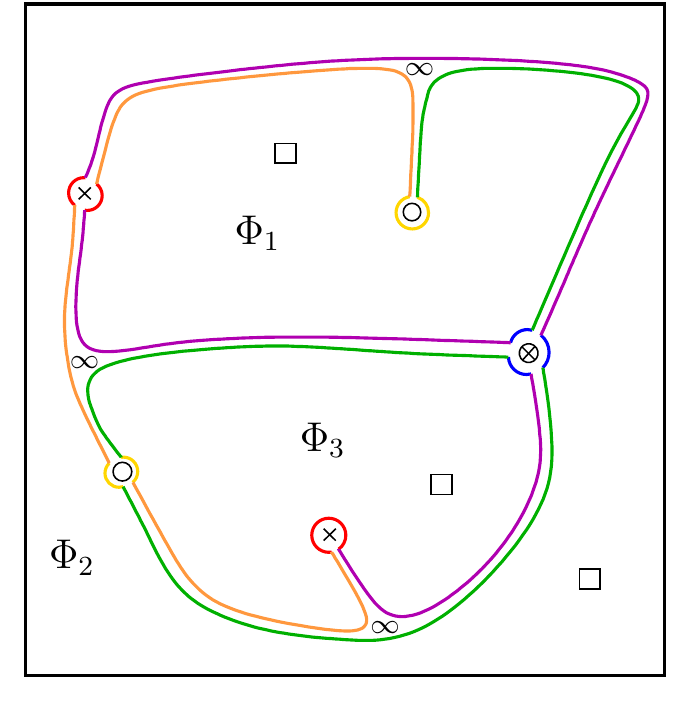}
   }
\caption{Diagram of the three-fold cover of the base space.  The inside of the ``cloverleaf'' in figure \ref{mapdiaga} is the simply connected patch, which has the three images shown in figure \ref{mapdiagb}.  The twist operators have 3 distinct indices: 1, 2, and 3.  In the base space, we have 3 functions $\Phi_i$, which in the covering space has been mapped to one function $\Phi$ on 3 patches.  We further show the three images of a generic point (where we will later include a non twist insertion), and the three images of infinity of the base space).   \label{mapdiag}}
\label{MapGraphic}
\end{figure}

To help visualize this better, we consider figure \ref{mapdiag} for the example of $\langle \sigma_{(12)} \sigma_{(23)} \sigma_{(321)}\rangle$.  We see the expanded images of the simply connected patch in figure \ref{mapdiagb}.  If we examine the ``lower island'' patch in figure \ref{mapdiagb}, we see that there is an isolated image of $\sigma_{(12)}$ in this patch (marked with an ``x'', surrounded by a red contour).  Since the function in this patch is not twisted by $\sigma_{(12)}$, this patch must be associated with $\Phi_3$.  We may consider the other patches similarly.

In this way, the Lunin-Mathur technique has mapped the problem of twisted boundary conditions to a problem in the covering space with normal boundary conditions.  Consider an arbitrary configuration of fields $\Phi_i$ in the base that satisfies the boundary conditions imposed by the twist operators.  We see that this must correspond to a unique configuration of $\Phi$ in the covering space.  The reverse is also true: every configuration on the covering surface corresponds to a configuration on the base space that satisfies the boundary conditions imposed by the twist fields.  Therefore, in calculating a correlator by integrating over all configurations $\Phi_i$ in the base space, one may instead compute a correlator in the covering space, integrating over all configurations $\Phi$: the path integrals are related.

It is crucial to account for the change to the measure of the path integral.  We must consider this because the locally conformal map used is not a member of the class of $SL(2,{\mathbb C})$ anomaly-free maps of the Riemann sphere, and the CFT we are dealing with has a non-zero central charge $c$.  The map induces a metric on the covering space $g$, which is scaled back to a reference metric $g=e^{\phi}g_c$.  The change to the measure by this rescaling of the metric is given by the exponential of the Liouville action \cite{Friedan:1982is}
\be
S_L=\frac{c}{96\pi} \int d^2t \sqrt{-g_c} \left[\pa_\mu \phi \pa_\nu \phi g^{\mu \nu}_c + 2 R(g_c) \phi\right].
\ee
This factor, coming from an anomaly, simply multiplies the correlators in the covering space,
\be
\langle \prod_i {\mathcal O}_i \rangle_{\rm base} = e^{S_L}\langle \prod_i \hat{\mathcal O}_i\rangle_{\rm cover}
\ee
To compute this factor, the Liouville action must be suitably regulated, as was done in \cite{Lunin:2000yv} to compute certain 3-point and 4-point functions.  When the insertions $\hat{\mathcal O}_i$ on the covering surface are set to $1$, this defines the ``bare twist'' correlation function.

For supersymmetric theories, one must consider lifting fermions to the cover as well \cite{Lunin:2001pw}.  In these cases, the conformal transformation properties of the of the fermions play an important role.  In the vicinity of a ramified point, a fermion field transforms as
\be
z=a t^n + \cdots \rightarrow \psi(t)=\left(\frac{dz}{dt}\right)^{\frac12} \psi(z)=\left(ant^{n-1}+\cdots\right)^{\frac12} \psi(z)\label{fermrephas}
\ee
where we consider the point to be at $z=0$ in the base space and $t=0$ in the covering space for simplicity.  To circle the point at $t=0$, we take $t\rightarrow \exp(2\pi i) t$, or in the base space, $z\rightarrow \exp(2\pi i n)$.  The twist operator is of order $n$, and so if we circle it $n$ times in the base space, the field comes back to itself.  Thus $\psi(\exp(2\pi i n)z)=\psi(z)$.  This means that in (\ref{fermrephas}), $\psi(\exp(2\pi i)t)=\exp(2\pi i (n-1)/2) \psi(t)$ in the covering space.  When $n$ is odd, $\psi(t)$ returns to itself.  However, if $n$ is even, $\psi(t)$ returns to minus itself.  In this case, $\psi(t)$ must be antiperiodic in the covering space to furnish a $\psi(z)$ in the base space that is periodic when circling $z=0$ $n$ times.  To account for this boundary condition, there must be a spin field $\mathcal{S}$ at the location of the ramified point in the covering space to ensure the correct periodicity conditions.

Finally, we wish to consider a large $N$ limit for applications in AdS/CFT.  It was shown in \cite{Lunin:2000yv} that, due to combinatoric factors, the leading order in $1/N$ is given by the case where the covering surface is a sphere.  Here, and in what follows, we will concentrate on these cases, although we believe that the techniques here and in the next section should extend to other Riemann surfaces.

\subsection{Generalization to the non twist sector.}
\label{gen}

The Lunin-Mathur technique was developed for twist sector operators, with twist sector excitations.  Here we will generalize the LM technology for non twist sector operators, and non-twist sector excitations of twist sector operators.

First we consider excitations of the twist operators by modes that are not twisted by the operator.  In such a case, the field acting on the twist operator shares no OPE with it, and so we can simply multiply the operators together.  For example, if we have a bare twist $\sigma_{(12)}$ and we wish to excite it with a mode of a bosonic $X$ operator, $\alpha_{3,-1}$ \footnote{The first subscript denotes the copy; the second denotes the mode.},
 we simply write this as $\pa X_3 \sigma_{(12)}$.  This is for the non-$S_N$-invariant operator.  The full $S_N$-invariant operator would involve summing over all images of the $S_N$ symmetry group, which now acts on all of the indices, $1,2,3...$ i.e.
\be
\tilde{\sigma}_{2}'= \pa X_3 \sigma_{(12)} + \pa X_1 \sigma_{(32)} +\pa X_4 \sigma_{(12)}+\pa X_1 \sigma_{(42)}+\cdots.
\ee
where we consider all $S_N$ permutations of the indices $1,2,3\cdots N$.  Half will be repeated operators of the same kind, because $\sigma_{(ij)}=\sigma_{(ji)}$.  We would like to figure out how to compute correlators with such excitations.  Again, it will be sufficient to consider only correlators of the non-$S_N$-invariant operators, and then sum to make an $S_N$-invariant correlator.

The prescription is as follows.  Imagine that we are considering a correlator involving twist operators, of which only one has an excitation in a copy that does not involve its twist indices.  When we expand out the gauge invariant operators, we will have two cases come up: either the twist directions of the other twist operator are along the direction of the excitation, or they are not.  For example, if we consider a 3-point correlator of $\tilde{\sigma}_{2}'$ along with itself and $\sigma_{3}$, there will be terms of the form
\be
\langle (\pa X_3 \sigma_{(12)}) (\pa X_1 \sigma_{(23)}) (\sigma_{(321)})\rangle\quad \mbox{ or } \quad  \langle (\pa X_4 \sigma_{(12)}) (\pa X_4 \sigma_{(23)}) (\sigma_{(321)})\rangle.
\ee
In the second case, the $\pa X_4$ terms factorize, because it does not have any directions in common with the twists in the fields.

However, in the first case, we see that $X_3$ and $X_1$ are directions that are associated with twist directions, just not directions for the operator that they act on to excite.  The solution to this problem is rather simple.  Recall that $\pa X_1$ adds a boundary condition for the field $X_1$.  The function $X_1$ is associated with a particular patch in the cover.  Thus, to add the excitation of this field, we make an insertion in the covering space in the patch associated with $X_1$.  The location of the insertion is at the image of the twist $\sigma_{(12)}$ in this patch.  In our diagram, \ref{mapdiagb}, we see that patch $\Phi_1$ ($X_1$ in our example) has a point associated with the twist operator $\sigma_{(12)}$ in the upper half of the diagram.  At this location, we make an insertion of
\be
\left. \left(\frac{dz}{dt}\right)^{-1}\right|_{t=t_{o,1}}\times\;\pa X(t_{o,1})
\ee
where $t_{o,1}$ is the image of the ``o'' point in the patch for $X_1$, i.e. the circle in \ref{mapdiagb} in the upper half of the diagram.  This produces the correct boundary condition for the field $X_1$ at the point marked with the ``o'' in the base space.  Similarly, we must make an insertion of
\be
\left. \left(\frac{dz}{dt}\right)^{-1}\right|_{t=t_{x,3}}\times\;\pa X(t_{x,3})
\ee
where $t_{x,3}$ is the image of the point marked with an ``x'' in patch for $X_3$, i.e. the ``x'' in the lower half of diagram \ref{mapdiagb}.  We would then need to add all other symmetric combinations as well.  This process can be generalized to more complicated excitations.

The general prescription is simple to state.  First, take the $S_N$-invariant operator, and expands this in terms of non-$S_N$-invariant pieces.  For each piece, see whether the twist parts of the operators agree to make sure that the product of all the cycles is $1$ in some order.  Note that $(12)(23)=(321)$ while $(23)(12)=(123)$, so $\sigma_{(12)}$ and $\sigma_{(23)}$ must fuse to both $\sigma_{(123)}$ and $\sigma_{(321)}$.  This was considered in \cite{Lunin:2000yv} when considering 4-point functions.  The basic observation was that if one takes $\sigma_{(123)}$ in a path around the $\sigma_{(12)}$ insertion, it becomes $\sigma_{(213)}=\sigma_{(321)}$.
Any excitations of these operators along directions not twisted by other operators simply factorize.  Any excitations along directions that become twisted by other operators in the correlator are accounted for by operator insertions at the appropriate image in the covering space.  Those parts of the operator that describe the excitations are mapped using the correct conformal transformation properties.  For example, a combination $\pa X \pa X$ would transform with an additional Schwarzian derivative piece when mapping to the cover, while $\pa \pa X$ would map as
\be
\pa^2 X(z)\rightarrow (\pa z/\pa t)^{-2}\pa^2 X(t)-(\pa z/\pa t)^{-3}(\pa^2 z/\pa t^2)\pa X(t).
\ee
To consider an arbitrary operator ${\mathcal{O}}\sigma_2$, where $\mathcal{O}$ describes some non-twist excitations, one only needs to know how $\mathcal{O}$ transforms under finite conformal transformations.

Next, we consider non twist insertions into the plane.  Let us illustrate this with another example.  Consider a simple type of non-twist sector field
\be
\mathcal{O}_0= \frac{1}{\sqrt{N}}\sum_{\kappa=1}^N \pa X_{\kappa}
\ee
where again we suppress all indices except for the copy index.  We may consider inserting such an operator in a correlator with twist sector fields.  The twist sector fields are made from sums over conjugacy classes of operators, as before.  These can be expanded into separate non-$S_N$-invariant contributions.  We want to address how to compute these non-$S_N$-invariant correlators individually, after which we can sum these together to find the correct $S_N$ invariant combination.

Without loss of generality, we may consider the twisted directions to be the first $s$ copies.  There are other combinations with similar operators involving $s$ copies of fields.  Some of these are just symmetric group images of the operators we are considering, and can be accounted for with a combinatoric factor, while others must be summed over.  Thus, we are considering a case where only the first $s$ fields have twisted boundary conditions.  Our non twist sector operator can then be written as the sum of two pieces
\bea
\mathcal{O}_0&=& \frac{1}{\sqrt{N}}\sum_{\kappa=1}^s \pa X_{\kappa} +\frac{1}{\sqrt{N}}\sum_{\kappa=s+1}^N\pa X_{\kappa} \nn \\
&=& \mathcal{O}_{0,\parallel} + \mathcal{O}_{0,\perp}
\eea
where $\mathcal{O}_{0,\parallel}$ is the first $s$ terms (copies along the twists), and $\mathcal{O}_{0,\perp}$ is the other terms (copies not along the twists).  All operators and excitations there of involving the $(s+1,...,N)$ copies appear in factorized correlators, and can be computed with extant LM technology.

This leaves us to compute a correlator involving $\mathcal{O}_{0,\parallel}$ and a set of twist operators with possible excitations along the first $s$ copies of the CFT, generically of the form
\be
A=\langle \mathcal{O}_{0,\parallel}(z_0) \prod_\ell \sigma_{n_\ell}(z_\ell)\rangle
\ee
where the twist fields $\sigma_{n_\ell}$ have twist indices along the first $s$ directions, and may have excitations along the first $s$ directions.

Now, we must lift the computation to the covering space, and come up with a covering space interpretation of $\mathcal{O}_{0,\parallel}(z_0)$.  We lift the excited twist operators as in the last example.  First note that the position of this operator $z_0$ is a generic point, and so it is uplifted to $s$ points in the cover $\{t_{0,i}\}$.  Each point on the cover is associated with a patch, as in diagram \ref{mapdiagb}.  Next, we note that the operator $\frac{1}{\sqrt{N}}\sum_{\kappa=1}^s \pa X_{\kappa}(z_0)$ has the interpretation as a sum of states.  Each of these states has boundary conditions on only one of the fields $X_1(z_0),..., X_s(z_0)$.  Since each of these fields is lifted to a particular patch in the cover, we see that we must make an operator insertion at only one of these points.  At which point must we make the insertion?  The answer is simple: we put the insertion at each image point $t_{0,i}$, and add the terms, just as the operator $\mathcal{O}_{0,\parallel}(z_0)$ is a sum of terms, each placing boundary conditions on different fields.  We lift $\pa X$ using its conformal transformation properties, i.e. in the cover the insertion at the $i^{\rm th}$ point is
\be
\left. \left(\frac{dz}{dt}\right)^{-1}\right|_{t=t_{0,i}}\times\;\pa X(t_{0,i}).
\ee
In our example in diagram \ref{mapdiag}, we would make the above insertions at one of the points marked with $\square$ in the covering space.  We would compute all three insertions, and then add the contributions together.

Again, we would generalize this the same way as above.  Take an operator $\mathcal{O}_{i_1\cdots i_q}$ that describes an non-$S_N$-invariant piece of an $S_N$-invariant operators $\sum_{S_N(i_k)}\mathcal{O}_{i_1\cdots i_q}$, which is in the non-twist sector.  Then, to lift this to the cover, we would need to put an operator insertion of various pieces of $\mathcal{O}$ in the cover, transforming each piece according to its finite conformal transformation properties.  For example, given a non-$S_N$-invariant operator $\pa X_1 \pa X_2$ (here $1,2$ are copy indices) we would put an image of $(\pa z/\pa t)^{-1}\pa X$ in patch 1 {\it and} in patch 2.  This is because the operator $\pa X_1 \pa X_2$ is associated with a state with an excitation in both the first $X_1$ mode, and the first $X_2$ mode.  This is just the same as in simple quantum mechanics: sums mean ``or'' while multiplication means ``and.''

Further, we can combine these techniques in a straightforward way.  We have given an interpretation for each of these kinds of operators in the covering space, and so combining them simply combines the steps.  We put appropriately transformed ``image operators'' at the appropriate images of the point in the covering space.  These can be found once the map is known.  This interpretation also sheds light on the meaning of the multiple images in the covering space, both for the twist sector operators when they appear with other twists, and also for the non-twist sector operators.

We have thus found how to compute non twist sector operators using a generalization to the Lunin-Mathur technique.  In the next section we will consider some example calculations, and show that results generated by this technique agree with the expected form for CFT correlators.

%--------+---------+---------+---------+---------+---------+---------+---------+

\section{Example Calculations}

\subsection{Excitations orthogonal to twist directions}
\label{ortho}

In the sections that follow, we show how to use our generalization of the LM technology to compute 3-point functions.  To make these computations, we will need the explicit form of the conformal maps.  Here, we restrict to the case where the covering surface is the two sphere, which corresponds to the leading order in $1/N=1/(N_1 N_5)$, as explained in \cite{Lunin:2000yv}.

The first map that we consider is for a correlator of the form $\langle \Sigma_2 \Sigma'_2 \Sigma_3\rangle$, involving two twist 2 operators, and a twist 3 operator, as shown in figure \ref{mapdiag}.  We consider the position of the operator insertions to be $z=a_1,a_2,\infty$ for the twist $2,2,3$ insertions respectively.   Some of the images of these points in the covering surface must be ramified.  Near these points int the cover, the map is locally $z-z_0=b(t-t_0)^n+\cdots$.  Different copies of the simply connected patch meet at ramified points in such a way as to give the correct boundary conditions for the fields.  For two twist 2 fields and one twist 3 field, the correct map to use is
\be
z-a_1=(a_1-a_2)t^2(2t-3)
\ee
which is correctly ramified at $t=0$ for a twist 2 operator.  Note that this allows us to find the other image of $z=0$ located at $t=3/2$.
We may consider the location of the other twist 2 operator, and see that
\be
z-a_2=(a_1-a_2)(2t+1)(t-1)^2
\ee
which is again correctly ramified at $t=1$ for a twist 2 operator.  We see the other image of $z=a_2$ is located at $t=-1/2$.  The third ramified point is clearly located at $t=\infty, z=\infty$, again with the correct behavior $z=2(a_1-a_2)t^3+\cdots$.

In what follows, it will be convenient to consider the twist operators at finite points.  We accomplish this with $SL(2,{\mathbb C})$ transformations in the $z$ and $t$ plane.  We map the locations of the twist insertions as follows: $z=a_1$ will map to $t=0$ as before, $z=a_2$ to $t=1$ as before, but now we will map the location of the twist three operator to be at $z=b$ and its image in the covering surface will be $t=\omega$.  Performing the needed $SL(2,{\mathbb C})$ transformations in the $z$ and $t$ planes leads to the map
\be
z-a_1=-\frac{t^2([1+2\omega]t-3\omega)(a_1-a_2)(a_1-b)(\omega-1)^2} {t^2([1+2\omega]t-3\omega)(a_1-a_2)(\omega-1)^2+(t-\omega)^3(a_2-b)}.
\ee
Using translation invariance of the base space we set $b=0$ and, using the $SL(2,\mathbb{C})$ invariance of the $t$ plane, we set $\omega=-1$.  This gives a simplified map
\be
z=-\frac{a_1 a_2 (t+1)^3} {4t^2(t-3)(a_1-a_2)-(t+1)^3a_2}.
\ee
which one can also write as
\bea
z-a_1=-\frac{4t^2(t-3)(a_1-a_2)a_1} {4t^2(t-3)(a_1-a_2)-(t+1)^3a_2} \label{neara1}\\
z-a_2=-\frac{(t-1)^2(5t+1)(a_1-a_2)a_2} {4t^2(t-3)(a_1-a_2)-(t+1)^3a_2} \label{neara2}
\eea
showing the correct ramifications at $(z=a_1, t=0), (z=a_2,t=1), (z=0,t=-1)$.

To determine the 3-point function for the bare twists, coming from the Liouville term, we may simply use the result of \cite{Lunin:2000yv}
\be
|C_{2,2,3}|^2=\frac{1}{3^{\frac{1}{12}c}2^{\frac{5}{9}c}} \label{fusioncoeff223}
\ee
and so for twist operators located at finite points, we have
\be
\langle \sigma_{(1,2)}(a_1)\sigma_{(2,3)}(a_2) \sigma_{(3,2,1)(b)}\rangle=\frac{|C_{2,2,3}|^2}
{|a_1-a_2|^{\frac{2c}{72}}|a_1-b|^\frac{2c}{9}|a_2-b|^\frac{2c}{9}}
\ee
where $c$ is the central charge of one copy of the CFT.

For an example, we will consider a setup where there is a free $X$ CFT as part of the full CFT (and then, of course, there are $N$ copies).  Recall that the order of the three twist operators are 2, 2, and 3.  Although this tells us what twist sector the operators are in, it does not tell us about the excitations.  To be specific, we consider the operator $\tilde{\sigma}_{2}'$ already discussed:
\be
\tilde{\sigma}_{2}'= \pa X_3 \sigma_{(12)} + \pa X_1 \sigma_{(32)} +\pa X_4 \sigma_{(12)}+\pa X_1 \sigma_{(42)}+\cdots.
\ee
Let us consider what happens when we sum over all permutations.  Each occurrence of $\sigma_{(12)}$ will get dressed with all possible $\pa X_i$ where $i$ is neither 1 nor 2. Thus, we may write out the operator as
\be
\tilde{\sigma}_{2}'=\sum_{\begin{subarray}{l}
        i,j=1 \\  i<j
      \end{subarray}}^N\sum_{\begin{subarray}{l}
        k=1 \\  k\neq i,j
      \end{subarray}}^N\pa X_k \sigma_{(ij)}.
\ee
To normalize, we note that there are $N-2$ terms in the sum over $k$, and $\binom{N}{2}$ terms in the sum over $i,j$.  This implies that we should normalize the operator with a factor
\be
\sigma_{2}'=\frac{1}{\sqrt{(N-2) {\binom{N}{2}}}}\sum_{\begin{subarray}{l}
        i,j=1 \\  i<j
      \end{subarray}}^N\sum_{\begin{subarray}{l}
        k=1 \\  k\neq i,j
      \end{subarray}}^N\pa X_k \sigma_{(ij)}.
\ee
The 3-point function we wish to consider is
\be
\langle \sigma_{2}'(a_1) \sigma_{2}'(a_2) \sigma_{3}(b)\rangle
\ee
where $\sigma_{3}$ is just a bare twist three operator, i.e.
\be
\sigma_3 = \frac{1}{\sqrt{2{\binom{N}{3}}}} \sum_{\mbox{3-cycles}} \sigma_{(i,j,k)}
\ee
and we set the location of the twist three operator to be $b=0$ using translation invariance.

Now we expand this in terms of non-$S_N$-invariant pieces.  First, we note that to have a nonzero answer for the 3-point function, we must have a combination in the twist sectors of $\sigma_{(i,j)}\sigma_{(j,k)}\sigma_{(i,j,k)}$ or $\sigma_{(i,j)}\sigma_{(j,k)}\sigma_{(k,j,i)}$.  These two possibilities give the same contribution, so we will simply account for them with a combinatoric factor of 2, and take the second possibility.  Without loss of generality, we may choose $i=1, j=2$, which brings in a combinatoric factor of $\binom{N}{2}$.  Then, there are $\binom{N-2}{1}$ ways to assign the last index $k$.  Thus, we find that
\bea
&& \langle \sigma_{2}'(a_1) \sigma_{2}'(a_2) \sigma_{3}(0)\rangle= \\
&& 2 \frac{\sqrt{(N-3)!3!}}{\sqrt{2}\sqrt{N!}}\left\langle\left[\sum_{\begin{subarray}{l}
        i=1 \\  i\neq 1,2 \end{subarray}}^N\pa X_i(a_1) \sigma_{(12)}(a_1)\right] \left[\sum_{\begin{subarray}{l} j=1 \\  j\neq 2,3
      \end{subarray}}^N\pa X_j(a_2) \sigma_{(23)}(a_2)\right]
      \left[\sigma_{(321)}(0)\right]\right\rangle\nn
\eea
We see that we can break up the sums into two parts.  The first part is where $i,j$ are not twisted by any operators in the correlator, i.e. $i,j\geq 4$.  The second part is where $i,j$ are twisted by operators in the correlator, $i=3, j=1$.  The terms for $i,j\geq 4$ result in a combinatoric factor of $N-3$, and so we find
\bea
&& \langle \sigma_{2}'(a_1) \sigma_{2}'(a_2) \sigma_{3}(0)\rangle= \\
&& 2 \frac{\sqrt{(N-3)!3!}}{\sqrt{2}\sqrt{N!}}\Bigg(\left\langle\left[\pa X_3(a_1)\sigma_{(12)}(a_1)\right]
\left[\pa X_1(a_2) \sigma_{(23)}(a_2)\right]
      \left[\sigma_{(321)}(0)\right]\right\rangle\nn \\
      &&\qquad + (N-3)\left\langle \sigma_{(12)}(a_1)\sigma_{(23)}(a_2)\sigma_{(321)}(0)\right\rangle\left\langle \pa X_4(a_1) \pa X_4(a_2) \right\rangle                          \Bigg)
\eea
The second term above clearly has the correct form of a 3-point function. It simply gives an additional factor of $1/(a_1-a_2)^2$, which is what should happen: the holomorphic weight $h$ of the twist-two operators have both increased by 1.  This affects the $1/(a_1-a_2)^{h_1 + h_2 -h_3}$ terms, but not the terms of the form $1/(a_1-0)^{h_1 - h_2 +h_3}$, nor any of the other antiholomorphic terms in the 3-point correlator.  We can simply add the weight of the $\pa X$ operator to that of the original operator because the $\pa X$ in question shares no OPE with the twist operator.

The other requires us to lift the computation to the covering surface.  We lift the computation as
\bea
&&\left\langle\left[\pa X_3(a_1)\sigma_{(12)}(a_1)\right]
\left[\pa X_1(a_2) \sigma_{(23)}(a_2)\right]
      \left[\sigma_{(321)}(0)\right]\right\rangle \\
      &&\rightarrow \left\langle \left[ \left.\left(\frac{\pa z}{\pa t}\right)^{-1}\right|_{t=3}\pa X(3)\right] \left[ \left.\left(\frac{\pa z}{\pa t}\right)^{-1}\right|_{t=-\frac{1}{5}}\pa X\left(-\frac{1}{5}\right)\right] \right\rangle
\eea
where we have plugged in the explicit locations of the other images of $a_1$ and $a_2$ given the map (\ref{neara1}), or equivalently (\ref{neara2}).  Here we use a notation $\rightarrow$ to mean ``lift to the cover, and strip the Liouville action''.  This is convenient and allows us to concentrate on the CFT calculation in the cover.  The above computation in the covering surface is trivial, and gives
\bea
\left\langle \left.\frac{\pa z}{\pa t}\right|_{t=3}\pa X(3) \left.\frac{\pa z}{\pa t}\right|_{t=-\frac15}\pa X\left(-\frac15\right) \right\rangle&=&\frac{16a_2}{9a_1(a_1-a_2)}\frac{16 a_1}{225 a_2 (a_1-a_2)}\frac{1}{(3-(-1/5))^2}\nn \\
&=&\frac{1}{81(a_1-a_2)^2}
\eea
This extra dressing, just like the orthogonal piece, changes the 3-point function to be of the proper form.  Combining everything, we find
\be
\langle \sigma_{2}'(a_1) \sigma_{2}'(a_2) \sigma_{3}(0)\rangle=2 \frac{\sqrt{(N-3)!3!}}{\sqrt{2}\sqrt{N!}} \frac{|C_{2,2,3}|^2\left(\frac{1}{81}+N-3\right)}{(a_1-a_2)_{\phantom{1}}^{\frac{c}{72}+2}a_1^{\frac{2c}{9}}a_2^{\frac{2c}{9}}
(\bar{a}_1-\bar{a}_2)_{\phantom{1}}^{\frac{c}{72}}\bar{a}_1^{\frac{2c}{9}}\bar{a}_2^{\frac{2c}{9}}}. \label{newthreepoint1}
\ee
One final concern is whether we have properly normalized the twist two operators.  The normalization would be accomplished by mapping the 2-point functions to the covering surface.  However, normalization is actually already taken care of in this case because
\bea
\langle \sigma_{2}'(1) \sigma_{2}'(0)\rangle &=& \frac{1}{(N-2){\binom{N}{2} }}\left\langle\sum_{\begin{subarray}{l}
        i,j=1 \\  i<j
      \end{subarray}}^N\sum_{\begin{subarray}{l}
        k=1 \\  k\neq i,j
      \end{subarray}}^N\pa X_k(1) \sigma_{(ij)}(1)
      \sum_{\begin{subarray}{l}
        i,j=1 \\  i<j
      \end{subarray}}^N\sum_{\begin{subarray}{l}
        k'=1 \\  k'\neq i',j'
      \end{subarray}}^N\pa X_{k'}(0) \sigma_{(i'j')}(0)\right\rangle \nn \\
      &=& \frac{1}{(N-2)}\left\langle\sum_{\begin{subarray}{l}
        k=3
      \end{subarray}}^N\pa X_k(1) \sigma_{(12)}(1)
      \sum_{\begin{subarray}{l}
        k'=3
      \end{subarray}}^N\pa X_{k'}(0) \sigma_{(12)}(0)\right\rangle \nn \\
      &=& \left\langle \pa X_3(1) \sigma_{(12)}(1)
      \pa X_{3}(0) \sigma_{(12)}(0)\right\rangle =\langle\sigma_{(12)}(1)\sigma_{(12)}(0) \rangle.
\eea
In other words, normalization of the operator is completely taken care of via the normalization of the bare twists.  This was accounted for in the original computations in \cite{Lunin:2000yv} which leads to (\ref{fusioncoeff223}), and so our result (\ref{newthreepoint1}) is indeed correctly normalized.

\subsection{Non-twist operator insertions}
\label{nontwist}

For our next 3-point point function, we will consider a slightly simpler map, but a more complicated field content.  We will be considering a correlation function of the type
\be
\langle \sigma_{0} \sigma_{2}'' \sigma_{2}''\rangle \label{genform022}
\ee
where $\sigma_{0}$ is a non twist insertion, and $\sigma_{2}''$ is a twist two sector field. The conformal map that we will use works for the case of only two twist operators, with twist order $n=2$.  The map for two twist-$n$ fields, putting them at $z=0$ and $z=\infty$, is
\be
z=b t^n.
\ee
If we are interested in putting the locations of the operators at finite points, we may use the map
\be
z=a \frac{t^n}{t^n-(t-1)^n}
\ee
where the location of the twist operators is now $z=0$ and $z=a$.  It is easy to check that
\be
z-a=a\frac{(t-1)^n}{t^n-(t-1)^n}
\ee
so the ramified points are at $t=0$ and $t=1$ in the covering space.

Next, we would like to consider a specific field theory for concreteness.  For this, we will use the D1-D5 CFT.  The moduli space of this CFT is conjectured to have an orbifold point, where the field content is that of a ${\mathcal{N}}=(4,4)$ CFT with $N=N_1 N_5$ copies.  Here, $N_1$ and $N_5$ denote the number of D1 and D5 branes respectively.  The field content for one copy is four real scalars $X^{i}$, and four real fermions in both the left and right moving sectors $\psi^{j}$, $\widetilde{\psi}^{k}$.

The presence of fermions complicates the lift to the covering surface when the twist is of even order.  We will be considering $n=2$, and so this is a concern for us.  At the location of the ramified points in the cover there are spin fields.  To deal with these spin fields, we will bosonize the fermions, and write the spin fields in terms of exponentials of the bosons.  In terms of these fields, we will have a total of six right moving fields $\phi_i(z)$ and six left moving fields $\widetilde{\phi}_i(\zb)$.  The first four of these fields will correspond to the original bosons in the theory, breaking the left and right moving parts into $\phi$ and $\widetilde{\phi}$.  The final two in each sector correspond to the bosonized fermions.   We will follow the notation of \cite{Lunin:2001pw} for the bosonized fields, and introduce the following vectors
\bea
A=(1,i,0,0,0,0), \qquad B=(0,0,1,i,0,0) \nn \\
c=(0,0,0,0,0,1), \qquad d=(0,0,0,0,1,0) \\
e=(0,0,0,0,1,-1), \qquad f=(0,0,0,0,1,1) \nn
\eea
Clearly $A\cdot \phi,A^*\cdot \phi,B\cdot \phi,B^*\cdot \phi$ form a complete basis for constructing fields associated with bosons, and $c\cdot \phi$ and $d\cdot \phi$ give a complete basis for discussing bosonized fermions.  The combinations $f\cdot \phi$ and $e\cdot \phi$ are what naturally appear in spin fields, while $A\cdot \phi$ and $B\cdot \phi$ are what transform naturally under the $SU(2)_1\times SU(2)_2=SO(4)$ internal symmetry of the four bosons.  Here, and in what follows, we will concentrate on the holomorphic sector of the theory.  The antiholomorphic sector will follow similarly, without further need for comment.

We can now explicitly state the non-twist insertion that we wish to consider.  We take, for simplicity,
\be
\mathcal{O}= \frac{1}{\sqrt{N}}\:A\cdot \sum_{i=1}^N\pa \phi_i \label{twist0field}
\ee
which is clearly invariant under the permutation group $S_N$, and is also properly normalized with respect to factors of $N$.

Note that here and in what follows we will ignore the effects of cocycles.  For the computations at hand, this should be sufficient, as we can explicitly construct cocycles such that the spin fields we will consider, $\exp(\pm i f \cdot \phi/2)$, have no additional operator dressing \cite{companion}.  Thus, there should not be any additional phases in the computations below.  For further details, see \cite{companion}.

Next, we will need to consider which operators in the twist 2 sector we wish to include.  We consider the left-moving part of a deformation operator and its conjugate, given in \cite{David:1999ec} as one of an exhaustive list of all (1,1) primary operators.  Writing this in the notation of \cite{Lunin:2001pw} in the covering space, we find
\bea
O_{A(12)}(z)\rightarrow \frac{1}{\sqrt{2} b^{5/8}}\left( : (A \cdot\pa \phi) e^{-i f\cdot \phi/2}: + : (B\cdot\pa \phi) e^{if\cdot\phi/2}\right): \label{defop1}\\
O_{A(12)}(z)^{\dagger}\rightarrow \frac{1}{\sqrt{2} b^{5/8}}\left(:(A^* \cdot \pa \phi) e^{i f\cdot \phi/2}: + :(B^*\cdot \pa \phi) e^{-if\cdot\phi/2}:\right). \label{defop2}
\eea
In this expression $b$ is the leading term in the expansion
\be
z=bt^2+\cdots.
\ee
We have used translation invariance in the base space to move the twist operator to $z=0$, and translation invariance in the covering space to have the location of the ramified point at $t=0$.  Both $O_{A(12)}$ and $O_{A(12)}^{\dagger}$ are both primary operators of weight $1$.

We wish to consider certain excitations of these fields in twist directions so that we get a nontrivial 3-point function.  We use the techniques of \cite{Lunin:2001pw} to excite this twist field using the bosons.  First, we note that in the neighborhood of $O_{A,(12)}$ (which we put at $z=0$ for convenience) that $\pa \phi_1^a$ and $\pa \phi_2^a$ for $a=1,2,3,4$ do not have well-defined periodicity conditions.  Instead, a more natural combination is $\pa \phi_1^a + \exp{(2\pi i m/2)} \pa \phi_2^a$.  When $m$ is odd, the field is antiperiodic, and when $m$ is even, it is periodic.  This is just decomposing the collection of fields $\phi_n$ into eigenvectors of the operation $1\rightarrow 2\rightarrow 1$.  From this, we naturally define modes of these operators as
\be
A^*\cdot \alpha^{(12)}_{-m/2}\equiv\oint \frac{dz}{2\pi i}z^{1-1-m/2} A^*\cdot\left(\pa {\phi}_1(z) +e^{2\pi i m/2} \pa \phi_2(z)\right) \label{excite}
\ee
where we can see that the integrand is single valued, and we have picked a certain direction for the excitation.  We lift the action of this current to the covering space as
\bea
[A^*\cdot\alpha^{(12)}_{-m/2}, O_{A,(12)}(z)^{\dagger}]&=&\oint \frac{dz}{2\pi i}z^{1-1-m/2} A^*\cdot\left(\pa {\phi}_1(z) + e^{2\pi i m/2} \pa \phi_2(z)\right) O_{A,(12)}(0)^{\dagger} \nn \\
&\rightarrow& \oint \frac{dt}{2\pi i}\left(\frac{dz}{dt}\right)^{1-1}(z(t))^{-m/2} (A^*\cdot \pa \phi(t))  \\
&& \times\frac{1}{\sqrt{2} b^{5/8}}\left(:(A^* \cdot \pa \phi) e^{i f/2\cdot \phi}(0): + :(B^*\cdot \pa \phi) e^{-if/2\cdot \phi}(0):\right). \nn
\eea
Note that $A^*$ is orthogonal to all vectors except $A$, and this does not appear anywhere in the twist operator under consideration.  Hence, there are no singular terms, and the above operator is normal ordered.  We simply need to expand $z(t)^{-m/2}$ and $\pa\phi(t)$ to the appropriate orders, and find the pole term.  It turns out that for the correlator that we consider later, $m=2$ is the first term that will give a nonzero 3-point function.  For $m=2$, we find
\bea
&& [A^*\cdot \alpha^{(12)}_{-2/2}, O_{A,(12)}^{\dagger}] \label{excitation2} \\ \label{defexcite}
&\rightarrow& \frac{b^{-1-5/8}}{\sqrt{2}} :\left(A^*\cdot\left( \pa^2 \phi-\frac{b_1}{b}\pa\phi\right)\right)\left((A^* \cdot \pa \phi) e^{i f/2\cdot \phi} + (B^*\cdot \pa \phi) e^{-if/2\cdot \phi}\right): \nn
\eea
where we have expanded the map to second order as
\be
z=bt^2+b_1t^3+b_2t^4+\cdots\label{mapexpandmost}
\ee

We have now constructed an operator that we know how to lift to the covering space.  However, we would like to check if this operator is a quasiprimary.  To do so, we will need to apply
\be
L_{\ell}=\oint \frac{dz}{2\pi i} z^{2-1+\ell}T(z).
\ee
To apply this in the covering space, recall that the stress tensor does not transform tensorially, but rather transforms as
\be
T(z)\rightarrow \left(\frac{dz}{dt}\right)^{-2}\left(T(t)-\frac{c}{12}\left\{z(t),t\right\}\right)
\ee
with $\{z(t),t\}$ denoting the Schwarzian derivative.  Lifting to the cover, the stress tensor is just
\be
T(z)=-\frac12 : \pa\phi^a \pa \phi^a:
\ee
and $c=6$ in the cover.  Thus, to check if our operator is a quasiprimary, we compute
\bea
&& [L_{\ell},[A^*\cdot\alpha^{(12)}_{-2/2}\;,\: O_{A,(12)}^{\dagger}]] \nn \\
&& \rightarrow \frac{b^{-1-5/8}}{\sqrt{2}}\oint \frac{dt}{2\pi i} \left(\frac{dz}{dt}\right)^{-1} (z(t))^{1+\ell} \left(-\frac12 :\pa \phi^{a} \pa \phi^{b}(t): \delta_{a b}-\frac12 \{z(t),t\}\right) \nn \\
&& \times :\left(A^*\cdot\left( \pa^2 \phi-\frac{b_1}{b}\pa\phi\right)\right)\left((A^* \cdot \pa \phi) e^{i f\cdot \phi/2} + (B^*\cdot \pa \phi) e^{-if\cdot \phi/2}\right)(0):.
\eea
for $\ell\geq 0$.  We note that the OPE we need to compute above starts at order $1/t^3$, due to the orthogonality of $A^*$ with itself and with $B^*$.  We must expand the functions of $t$ according to (\ref{mapexpandmost}), and we find
\bea
\left(\frac{dz}{dt}\right)^{-1}=\frac{1}{2bt}\left(1-\frac{3b_1}{2b}t-\frac{8b_2b-9b_1^2}{4b^2}t^2 + \cdots\right),
\eea
and
\bea
(z(t))^{1+\ell}=b^{1+\ell}t^{2+2\ell}\left(1+\frac{(1+\ell)b_1}{b}t+\frac{(1+\ell)(2b_2b+b_1^2\ell)}{2b^2}t^2+\cdots\right).
\eea
Therefore, the leading order behavior of these terms put together is $t^{1+2\ell}$.  Putting this together, we expand out
\bea
&& [L_{\ell},[A^*\cdot\alpha^{(12)}_{-2/2}, O_{A,(12)}^{\dagger}]] \nn \\
&&\rightarrow \frac{b^{-1-5/8+\ell}}{2\sqrt{2}}\oint \frac{dt}{2\pi i} t^{1+2\ell}\left(1-\frac{3b_1}{2b}t+\cdots \right)\left(1+\frac{(1+\ell)b_1}{b}t+\cdots \right)\nn \\
&& \times \left(-\frac12 :\pa \phi^{a} \pa \phi^{b}(t): \delta_{a b}+\frac{3}{4t^2}+\cdots\right) \nn \\
&& \times :\left(A^*\cdot\left( \pa^2 \phi-\frac{b_1}{b}\pa\phi\right)\right)\left((A^* \cdot \pa \phi) e^{i f/2\cdot \phi} + (B^*\cdot \pa \phi) e^{-if/2\cdot \phi}\right)(0):.
\eea
We can xpand the OPE as
\bea
&&-\frac12 \delta_{a b} :\pa \phi^{a} \pa \phi^{b}(t):\;\; :\left(A^*\cdot\left( \pa^2 \phi-\frac{b_1}{b}\pa\phi\right)\right)\left((A^* \cdot \pa \phi) e^{i f/2\cdot \phi} + (B^*\cdot \pa \phi) e^{-if/2\cdot \phi}\right): \nn \\
&&= \frac{2}{t^3}:A^*\pa \phi e^{if/2\cdot \phi}\left((A^* \cdot \pa \phi) e^{i f/2\cdot \phi} + (B^*\cdot \pa \phi) e^{-if/2\cdot \phi}\right): \nn \\
&&+ \frac{1}{t^2}:\left[A^*\cdot\left(\left[3+\frac{(f/2)^2}{2}\right]\pa^2 \phi-\left[2+\frac{(f/2)^2}{2}\right]\frac{b_1}{b}\pa\phi\right)\right] \nn \\
&& \qquad \qquad \qquad \qquad \times \left[(A^* \cdot \pa \phi) e^{i f/2\cdot \phi} + (B^*\cdot \pa \phi) e^{-if/2\cdot \phi}\right]: \nn \\
&&+\frac{1}{t}\pa\left[\left(A^*\cdot\left( \pa^2 \phi-\frac{b_1}{b}\pa\phi\right)\right)\left((A^* \cdot \pa \phi) e^{i f/2\cdot \phi} + (B^*\cdot \pa \phi) e^{-if/2\cdot \phi}\right)\right]+ \cdots
\eea
with all operators at the location $t=0$.  Combining the leading order behavior of the prefactors $t^{1+2\ell}$ with the leading order singularity of the OPE $1/t^3$, we see that the leading in the integral is $t^{-2+2\ell}$.  We can see that the $\ell\geq1$ will give $0$ in the contour, so that this operator is not just quasiprimary, but is in fact (Virasoro) primary, as long as it has a well defined conformal dimension.  To determine this, we must consider $\ell=0$.  For this value of $\ell$, we see that the $3/(4t^2)$ contribution from the Schwarzian must be taken into account.  Specializing to $\ell$=0, we find
\bea
&& [L_{0},[A^*\cdot\alpha^{(12)}_{-2/2}, O_{A,(12)}^{\dagger}]] \nn \\
&& \rightarrow -\frac{b^{-1-5/8}}{2\sqrt{2}} \frac{b_1}{b}:A^*\pa \phi e^{if\cdot \phi/2}\left((A^* \cdot \pa \phi) e^{i f\cdot \phi/2} + (B^*\cdot \pa \phi) e^{-if\cdot \phi/2}\right):  \\
&& + \frac{b^{-1-5/8}}{2\sqrt{2}}:\left[A^*\cdot\left(\left[3+\frac{(f/2)^2}{2}+3/4\right]\pa^2 \phi-\left[2+\frac{(f/2)^2}{2}+3/4\right]\frac{b_1}{b}\pa\phi\right)\right] \nn \\
&& \qquad \qquad \qquad \qquad \times \left[(A^* \cdot \pa \phi) e^{i f\cdot \phi/2} + (B^*\cdot \pa \phi) e^{-if\cdot \phi/2}\right]: \nn\\
&=& \frac{\left[3+\frac{(f/2)^2}{2}+3/4\right]}{2}\frac{b^{-1-5/8}}{\sqrt{2}}:\left[A^*\cdot\left(\pa^2 \phi-\frac{b_1}{b}\pa\phi\right)\right] \nn \\
&& \qquad \qquad \qquad \qquad \times \left[(A^* \cdot \pa \phi) e^{i f\cdot \phi/2} + (B^*\cdot \pa \phi) e^{-if\cdot \phi/2}\right]: \nn\\
\eea
and so the $b_1/b$ terms have conspired to give us back the same operator, compare (\ref{excitation2}).  Plugging in $f^2=2$, we read off the conformal weight of this operator as $h=2$.  This is expected for a weight $1$ excitation working on a weight $1$ field.  Further, this shows that this operator is in fact a Virasoro primary operator.

Finally, we are ready to compute the 3-point function
\be
\frac{1}{\sqrt{N}}\frac{1}{{\binom{N}{2}}}\left\langle \big( A\cdot\left(\pa \phi_1(a)+\pa \phi_2(a) +\cdots\right)\big)\;\left(O_{A,(12)}(b)+\cdots\right)\;\left([\alpha^{(12)}_{-m/2}, O_{A,(12)}(0)^{\dagger}]+\cdots\right)\right\rangle \label{3pointcomponent}
\ee
where $\cdots$ denotes all of the other terms that lead to permutation invariant operators.  The extra factors of $1/\sqrt{\binom{N}{2}}$ normalize the sum of twist two operators $O_{A,(i,j)}$ with respect to $N$.  Above, there are $\binom{N}{2}$ total terms for the twist $2$ sector operators.  We will find nonzero expectation values only when the total twist is zero.  Hence, we would calculate $\binom{N}{2}$ terms, all of which give the same contribution, canceling the normalization factors coming from the twist two operators.  We find
\be
\frac{1}{\sqrt{N}}\left\langle \Big(A\cdot\left(\pa \phi_1(a)+\pa \phi_2(a) \right)\Big)\; O_{A,(12)}(b)\; \Big[\alpha^{(12)}_{-m/2}, O_{A,(12)}(0)^{\dagger}\Big]\right \rangle.
\ee
We have dropped the higher terms in the sum $\pa \phi_1 +\pa \phi_2 + \cdots$ because these factorize, and do not give contributions.  Note that the order of this interaction is $1/\sqrt{N}$, which is expected on general grounds from \cite{Lunin:2000yv}.

To lift this computation to the cover, as mentioned, we will need the map
\be
z(t)=b\frac{t^2}{t^2-(t-1)^2}=b\frac{t^2}{2t-1}. \label{twotwistmap}
\ee
This maps the first twist operator (at $b$) to $t=1$, and the excited twist operator to $t=0$.  For future reference, we expand around these points to find
\be
z(t)=-b\left(t^2+2t^3+\cdots\right), \qquad z(t)-b=b\Bigg((t-1)^2-2(t-1)^3+\cdots \Bigg) \label{map2expand}
\ee
We need to find the two images of $a$ in the covering space, which we call $t_\pm$.  These are determined from the map, and so we solve
\be
a=b\frac{t_{\pm}^2}{2t_{\pm}-1} \label{solvetpm}
\ee
to find
\be
t_\pm = \frac{a\pm \sqrt{a(a-b)}}{b}. \label{doubleimage}
\ee
Further, because the operator $\pa \phi_1(a)$ transforms under conformal mapping, we will need to compute
\be
\left.\frac{\pa z}{\pa t}\right|_{t=t_{\pm}} =\frac{2bt_{\pm}(t_\pm-1)}{(2t_\pm-1)^2}=\frac{2a(a-b)}{b t_\pm (t_\pm-1)}
\ee
where in the second equality we have used the definition (\ref{solvetpm}) to remove the terms $(2t_{\pm}-1)^2$.  We now lift the 3-point function (\ref{3pointcomponent}) to the covering space, to find
\be
\left \langle\Big( A\cdot\left(\pa \phi_1(a)+\pa \phi_2(a)\right)\Big)\;O_{A,(12)}(b)\;\Big[\alpha^{(12)}_{-2/2}, O_{A,(12)}(0)^{\dagger}\Big]\right \rangle.\label{3pointmis2}
\ee
We lift this computation to the covering surface, needing two images of the non twist operator, which we then add together.  This gives
\bea
&&\kern-3em \left \langle\Big( A\cdot\left(\pa \phi_1(a)+\pa \phi_2(a)\right)\Big)\;O_{A,(12)}(b)\;\Big[\alpha^{(12)}_{-2/2}, O_{A,(12)}(0)^{\dagger}\Big]\right \rangle\nn \\
&& \kern-3em\rightarrow \Bigg\langle \left[A\cdot\left(\frac{\pa z}{\pa t}(t_+)\pa \phi_1(t_+)+\frac{\pa z}{\pa t}(t_-)\pa \phi_1(t_-)\right)\right]\nn \\
&&\quad \times \frac{1}{\sqrt{2} b^{5/8}}\bigg[(A \cdot\pa \phi) e^{-i f\cdot \phi/2} + (B\cdot\pa \phi) e^{if\cdot \phi/2}\bigg](1) \\
&&\quad \times\frac{(-b)^{-1-5/8}}{\sqrt{2}}\bigg[A^*\cdot \left(\pa^2 \phi-\frac{b_1}{b} \pa \phi\right)\left(A^* \cdot \pa \phi\: e^{i f\cdot \phi/2} + B^*\cdot \pa \phi\: e^{-if\cdot \phi/2}\right)\bigg](0)\Bigg\rangle.\nn
\eea
We see from (\ref{map2expand}) that $b_1/b=2$.  Finally, we must match factors of $f\cdot \phi$ in exponents, and we get only two contributions
\bea
&&=\frac{(-1)^{-13/8}}{b^{1+5/4}} \sum_{\pm}\frac{b t_\pm (t_\pm-1)}{2a (a-b)} \\
&& \qquad \qquad\times\Bigg[ \bigg\langle A\cdot\pa\phi(t_\pm)[A\cdot\pa \phi e^{-if\cdot \phi/2}(1)][A^*\cdot(\pa^2\phi-2\pa\phi)A^*\cdot\pa\phi e^{if\cdot\phi/2}(0)]\bigg\rangle \nn \\
&&\qquad \qquad \qquad +\bigg\langle A\cdot\pa\phi(t_\pm)[B\cdot\pa \phi e^{if\cdot \phi/2}(1)][A^*\cdot(\pa^2\phi-2\pa\phi)B^*\cdot\pa\phi e^{-if\cdot\phi/2}(0)]\bigg\rangle\Bigg].\nn
\eea
We see that the first expectation value has two types of contractions to remove $\pa\phi^a$, while the second has only one.  These become
\bea
 && =\frac{(-1)^{-13/8}}{b^{1+5/4}}\sum_{\pm}\frac{b t_\pm (t_\pm-1)}{2a (a-b)}  \\
 &&\qquad \qquad \times \Bigg[ (A^*\cdot A)\left(\frac{-2}{t_{\pm}^3}-2\frac{-1}{t_{\pm}^2}\right)(A^*\cdot A)\frac{-1}{(1)^2}\langle e^{-if\cdot \phi/2}(1)e^{i f\cdot \phi/2}(0) \rangle\nn \\
&&\qquad \qquad \qquad +(A^*\cdot A)\left(\frac{-1}{t_{\pm}^2}\right)(A^*\cdot A)\left(\frac{-2}{1^3}-2\frac{-1}{1^2}\right)\langle e^{-if\cdot \phi/2}(1)e^{i f\cdot \phi/2}(0) \rangle\nn \\
&& \qquad \qquad \qquad + (A^*\cdot A)\left(\frac{-2}{t_{\pm}^3}-2\frac{-1}{t_\pm^2}\right)(B^*\cdot B)\frac{-1}{1^2}\langle e^{if\cdot \phi/2}(1)e^{-i f\cdot \phi/2}(0) \rangle\Bigg].\nn
\eea
Note that the second line above gives no contribution.  Further, there are several combinations of the form
\be
\frac{t_\pm(t_\pm-1)}{t_\pm^2}=\frac{(t_\pm -1)}{t_\pm}.
\ee
Explicitly summing these contributions using (\ref{doubleimage}), we find
\be
\frac{(t_+-1)}{t_+}+\frac{(t_- -1)}{t_-}=0.
\ee
This type of cancelation is what causes the $m=1$ case to vanish, and is the reason that we did not use this seemingly simpler calculation to illustrate our technique.  Summing the other terms gives the result
\be
=\frac{4(-1)^{-13/8}}{b^{5/4}a(a-b)}\left(\frac{t_+-1}{t_+^2}+\frac{t_--1}{t_-^2}\right).
\ee
Evaluating the sum, after plugging in $t_+$ and $t_-$, we find that the 3-point function (\ref{3pointmis2}) lifts to the remarkably simple answer
\bea
\left\langle \bigg(A\cdot\left(\pa \phi_1(a)+\pa \phi_2(a)\right)\bigg)\: O_{A,(12)}(b)\: \bigg[\alpha^{(12)}_{-2/2}, O_{A,(12)}(0)^{\dagger}\bigg]\right\rangle \nn \\
\rightarrow \frac{-8(-1)^{-13/8}}{b^{5/4}a^2}.
\eea
Of course this is only the lifted part of the computation.  We must include the contribution from the Liouville term as well.  We read this from equation (3.18) of \cite{Lunin:2000yv}, recalling that the bare twists have weight $h=(c/24)(n-1/n)=3/8$ for $c=6,n=2$.  Hence, we must dress the above computation with a factor of $b^{-3/4}$.  Our total 3 point function becomes
\bea
&&\langle A\cdot\left(\pa \phi_1(a)+\pa \phi_2(a)\right)O_{A,(12)}(b)[\alpha^{(12)}_{-2/2}, O_{A,(12)}(0)^{\dagger}]\rangle \nn \\
&&\propto \frac{-8(-1)^{-13/8}}{b^{2}a^2}
\eea
up to the normalization factors from \cite{Lunin:2000yv} which we have not included, and normalizations of (\ref{twist0field}), (\ref{defop1}) and (\ref{defexcite}), which we could compute by calculating the 2-point functions.  To faithfully include the normalization from \cite{Lunin:2000yv}, we would need to include what was happening in the right moving sector as well, include the extra factors of $\bar{b}$ from the right moving spin fields in the covering space.  This would help cancel some of the phase ambiguity in $(-1)^{-1-5/8}$ because there would be a factor of $(-\bar{b})^{-5/8}$: we would take these phases to be opposite in direction and cancel to give an overall $-1$ factor.  However, what is important for us here is that the above expression exactly matches the expected behavior for a 3-point function of quasi primary fields:
\be
\langle {\mathcal A}(a){\mathcal B}(b){\mathcal C}(0)\rangle = \frac{C_{\mathcal A B C}}{(a-b)^{h_{\mc A}+h_{\mc B}-h_{\mc C}} a^{h_{\mc A}+h_{\mc C}-h_{\mc B}}b^{h_{\mc A}+h_{\mc B}-h_{\mc C}}}=\frac{C_{\mathcal A B C}}{a^2b^2}
\ee
where $h_{\mc A}=1,h_{\mc B}=1,h_{\mc C}=2$.  Note that this behavior comes about only {\it after} we summed over different images.  Each term had contributions from the conformal transformation properties of $\pa \phi$ and from the particular images $t_\pm$ that get mapped to the postion $z=a$.  These all come together to give a result that is meaningful in the base space.

\section{Discussion}
\label{disc}
We now comment on our generalization of the LM covering space technique, some of its features, and possible obstacles to overcome.  First, we note that in the generalization to the non twist sector operators we could compute the images of the non-twist field in the covering space explicitly.  However, for general maps, where more covers of the space are required, we expect this straightforward approach to begin to run into difficulties.  The relative simplicity we observed was partially aided by the low order of the twists, but mainly came from the low number of twist operators.  If we had instead considered twist-$n$ fields, the corresponding map with the twists at $z=0,t=0$ and $z=\infty, t=\infty$ would be
\be
z=t^n.
\ee
In this context, finding the images in the covering space is trivial.  We would just find the $n^{\rm th}$ roots of the location of the non-twist insertion $z=a$.  To find the images when the operators are at finite positions, one just feeds the various values of $a^{1/n}$ through the $SL(2,{\mathbb C})$ maps.  The real difficulty would come in when dealing with more complicated maps involving multiple covers and additional twist fields.  For a map involving $s$ covers, one would need to solve an $s^{\rm th}$ order polynomial in order to deal with the positions explicitly.  This is obviously impossible in general.  However, we do not need to know the individual pieces, we only need the sum.  One can imagine more sophisticated methods being available to obtain results in these cases, relating sums of powers of solutions to polynomial equations to various coefficients of the polynomial.  This seems to be the only way for the summation to make sense in the base space.  This problem should also be important for the generalization to non-twisted excitations as well, given that the locations of the non-ramified images of the twist operator insertions are found by solving polynomials of an appropriately lower order.

One issue that our examples did not deal with is the possibility that there are different maps that give the correct ramifications in the covering space.  A problem related to this multiplicity of maps was recently considered in \cite{Pakman:2009ab}.  It would be interesting to see how to incorporate information about different maps {\it and} different images into the same formulation.  For example, seeing whether summing over images in the covering space yields sensible results in the base space for every map, or whether one must combine the information in different maps to make a sensible result, or whether working at large $N$ limits the possibilities.  If we consider the possibility that there are different maps, we note that each of these maps will give a different way to sew the multiple images of the simply connected patch together.  These different ways to sew together will give different ``topologies'' of field configurations that satisfy the correct boundary conditions.  In this way, the path integral may factorize into these distinct topologies.  It could be that the result found in \cite{Pakman:2009ab} is some statement about summing over these distinct topologies, each of which contributes once, but because of the special nature of the extremal correlators, the contributions from each topology is the same.   Further, the presence of multiple maps may be related to the ambiguity in the order that various group elements are multiplied to get the identity.  Of course, all of these considerations may be more a statement about how one is computing non-$S_N$-invariant pieces, and then needing to sum results to make $S_N$-invariant correlators.  Exploring these type of calculation should shed more light on the general process.  We will leave these questions to future work.

There is one well known system where the above techniques {\it can} be used unmolested: moving away from the orbifold point of the D1-D5 CFT.  For this, the theory needs to be deformed, and the correct operator to add to the action lives in the twist-2 sector of the theory.  Super chiral primary operators are protected from perturbative changes to their conformal dimensions.  Some of these are light operators, and correspond to supergravity modes, and some of these supergravity modes are in the non-twisted sector of the CFT.  For this reason, one may not simply ignore the interactions between the twist and non-twist sectors, and hope that it becomes unimportant in the gravity limit.  Further, we would like to be able to track what happens to these modes when the perturbation is turned on, as a start on bridging the gap between strong and weak coupling.  We will use the techniques developed here to begin to address these issues in a companion work \cite{companion}.

\section*{Acknowledgements}
The authors wish to thank Samir Mathur for ideas, interesting discussions, and guidance, and for hospitality during IGZ's visits.  We also wish to thank Steven Avery for helpful discussions.

AWP wishes to thank KITP at UCSB for support during the ``Bits `n' Branes" workshop. IGZ is grateful to the Simons Center for Geometry and Physics for support during the {``Superconformal Theories in Diverse Dimensions"} workshop.

This research was supported by the Canadian Institute of Particle Physics (IPP) and the Natural Sciences and Engineering Research Council (NSERC) of Canada.

%--------+---------+---------+---------+---------+---------+---------+---------+

\end{document}